\documentclass[12pt]{iopart}
\usepackage[utf8]{inputenc}% Some other (several out of many) possibilities
\usepackage{graphicx}% Include figure files
\usepackage{graphics}
\usepackage{dcolumn}% Align table columns on decimal point
\usepackage{bm}% bold math
\usepackage{float}
\usepackage[labelsep=space]{caption}

\begin{document}

\title[Constraints on light neutrino parameters derived from the study of $0\nu\beta\beta$ decay]{Constraints on light neutrino parameters derived from the study of neutrinoless double beta decay}

\author{Sabin Stoica$^1$, Andrei Neacsu$^2$}
\address{Horia Hulubei Foundation, P.O. Box MG-12 and \\Horia Hulubei National Institute of Physics and Nuclear Engineering, \\P.O. Box MG-6, Magurele-Bucharest 077125, Romania}
\ead{$^1$stoica@theory.nipne.ro, $^2$nandrei@theory.nipne.ro}

\date{\today}% It is always \today, today,

\date{\today}% It is always \today, today,

														
%\keywords{neutrino properties, double beta decay, neutrino mass}%Use showkeys class option if keyword display desired

\begin{abstract}
\noindent
The study of the neutrinoless double beta ($0\nu\beta\beta$) decay mode can provide us with important information on the neutrino properties, particularly on the electron neutrino absolute mass.   
In this work we revise the present constraints on the neutrino mass parameters derived from the $0\nu\beta\beta$ decay analysis of the 
 experimentally interesting nuclei. We use the latest results for the phase space factors (PSFs) and nuclear matrix elements (NMEs), 
 as well as for the experimental lifetimes limits. For the PSFs we use values computed with an improved method reported very recently. 
 For the NMEs we use values chosen from literature on a case-by-case basis, taking advantage of the consensus reached by the community on several 
 nuclear ingredients used in their calculation. Thus, we try to restrict the range of spread of the NME values calculated with different methods and, 
 hence, to reduce the uncertainty in deriving limits for the Majorana neutrino mass parameter. Our results may be useful to have an up-date image on the 
 present neutrino mass sensitivities associated with $0\nu\beta\beta$ measurements for different isotopes and to better estimate the range of values of the 
 neutrino masses that can be explored in the future double beta decay (DBD) experiments.
\end{abstract}

\maketitle
\renewcommand{\arraystretch}{1.2}

\section{Introduction}
Neutrinoless double beta decay is a beyond Standard Model(BSM) process by which an even-even nucleus transforms into another even-even nucleus 
with the emission of two electrons/positrons but no antineutrinos/neutrinos in the final states. Its study is very attractive because it would 
clarify the question about the lepton number conservation, decide on the neutrinos character (are they Dirac or Majorana particles?) and give 
a hint on the scale of their absolute masses. Moreover, the study of the $o\nu\beta\beta$ decay has a broader potential to search for other 
BSM phenomena. The reader can find up-to-date information on these studies from several recent reviews \cite{AEE08}-\cite{VOG12}, which also 
contain therein a comprehensive list of references in the domain. 

The scale of the absolute mass of neutrinos is a key issue for understanding the neutrino properties. It cannot be derived from neutrino 
oscillation experiments which can only measure the square of the neutrino  mass differences between different flavors 
\cite{KAE10}-\cite{KML11}. Analysis of $0\nu\beta\beta$ decay and cosmological data are at present the most sensitive ways to investigate this issue.    

 The lifetimes of the $0\nu\beta\beta$ decay modes can be expressed, in a good approximation, as a product of a phase space factor  
 (depending on the atomic charge and energy released in the decay, $Q_{\beta\beta}$), a nuclear matrix element (related to the nuclear structure of the 
 parent and daughter nuclei), and a lepton number violation (LNV) parameter (related to the BSM mechanism considered). 
 Thus, to extract reliable limits for the LNV parameters we need accurate calculations of both PSFs and NMEs, as well as reliable measurements of the lifetimes. 

The largest uncertainties in theoretical calculations for DBD are related to the NMEs values. That is why there is a continuous effort in literature to 
develop improved nuclear structure methods for their computation. At present, the NMEs are computed by several methods which differ conceptually, 
the most employed being the  proton-neutron Quasi Random Phase Approximation (pnQRPA) \cite{ROD07}-\cite{SK01}, Interacting Shell Model(ISM) \cite{Cau95}-\cite{HS10}, 
Interacting Boson Approximation (IBA) \cite{BI09}-\cite{iba-prc2013}, Projected Hartree Fock Bogoliubov (PHFB) \cite{RAH10} and Energy Density Functional (EDS) 
method \cite{RMP10}. There are still large differences %in the literature 
between the NMEs values computed with different methods and by different groups, and 
these have been largely discussed in the literature 
(see for example \cite{FAE12}-\cite{VOG12}). On the other side, there is a consensus in the community on the way that several nuclear effects and nuclear 
parameters should be used in calculations. In this work, we take advantage of this consensus when we chose the NMEs values, trying to restrict their range 
of spread, and consequently, to reduce the uncertainty in deriving the neutrino Majorana mass parameters.

Unlike the NMEs, the PSFs have been calculated a long time ago \cite{PR59}-\cite{SC98} and were considered to be computed with enough precision. 
However, recently, they were recalculated within an improved approach, by using exact electron Dirac wave functions (w.f.) taking into account the 
finite nuclear size and electron screening effects \cite{ki-prc85}. The authors found differences between their results and those calculated previously 
with approximate electron w.f., especially for heavier nuclei. We have also independently recalculated the PSFs by developing new routines for computing 
the relativistic (Dirac) electron w.f. by taking into account the nuclear finite size and screening effects. In addition, we use a 
Coulomb potential derived from a realistic proton density distribution in the daughter nucleus \cite{stoica-mirea-prc88}-\cite{MNPS13-RRP}. 
In this work we use new PSFs values obtained by improving the numerical precision of our routines as compared with our previous works. 
The obtained values are very close to those reported in refs. \cite{ki-prc85}-\cite{stoica-mirea-prc88}. 
 
Finally, for the lifetimes limits, we take the most recent results found in literature.

The paper is organized as follows. In the next section we shortly remind the general formalism for the derivation of the neutrino mass parameters 
from $0\nu\beta\beta$ decay analysis, highlighting the nuclear ingredients involved in calculations. In Section 3 we discuss the way of choosing the NME 
values and report our results for the light neutrino Majorana mass parameters, while in Section 4 we formulate the conclusions of our work.

\section{Formalism}

 We shortly present the general formalism for the derivation of neutrino mass parameters from $0\nu\beta\beta$ decay analysis. 
 We start with the lifetime formula and then describe the main steps and ingredients used in the theoretical calculation of their components, i.e. PSFs and NMEs.   

Assuming that the dominant mechanism of occurrence for the $0\nu\beta\beta$ decay mode is the exchange of Majorana 
left-handed light neutrinos between two nucleons from the parent nucleus, the lifetime reads:  

\begin{equation}
\left( T^{0\nu}_{1/2} \right)^{-1}=G^{0\nu}(Q_{\beta\beta}, Z)\mid M^{0\nu}\mid^2 \left(\frac{\left<m_{\nu}\right>}{m_e}\right)^2 \ ,
\end{equation}
%\noident
where $G^{0\nu}$ are the PSFs for this decay mode, depending on the energy decay $Q_{\beta\beta}$ and nuclear charge 
$Z$, $M^{0\nu}$ are the corresponding NMEs, depending on the nuclear structure of the parent and daughter nuclei involved in 
the decay, $m_e$ is electron mass and $\left<m_{\nu}\right>$ is the light neutrino Majorana mass parameter. This parameter can be expressed as a (coherent) linear combination of the light neutrino masses:

\begin{equation}
\left<m_{\nu}\right> = \mid \sum_{k=1}^3 U_{ek}^2 m_k\mid
\end{equation} 
\noindent
where $U_{ek}$ are the elements of the first row in the PMNS (Pontecorvo-Maki-Nakagawa-Sakata) neutrino matrix and $m_k$ are the light neutrino masses \cite{KL13}. 
From Eq. (1) the expression of $m_\nu$ reads:

\begin{equation}
\left<m_{\nu}\right> = \frac{m_e}{\mid M^{0\nu}\mid \sqrt{T^{0\nu}\cdot G^{0\nu}}}  
\end{equation}
\noindent

For deriving $\left<m_{\nu}\right>$ we need accurate calculations of both PSFs and NMEs, for each isotope for which there are experimental 
lifetime limits. The PSFs have been calculated a long time ago in some approximations \cite{PR59}-\cite{SC98} and were considered, until recently, 
to be computed with enough precision. However, they were recalculated recently in refs. \cite{ki-prc85}-\cite{MNPS13-RRP} using more advanced 
approaches for the numerical evaluation of the Dirac wave functions with the inclusion of nuclear finite size and screening 
effects. In addition, in ref. \cite{stoica-mirea-prc88} the usual Coulomb spherical potential was replaced by another one, 
derived from a more realistic proton density distribution in the daughter nucleus. 
These recent PSF calculations led to significant differences in the comparison to the older calculations, 
especially for the heavier isotopes, that should be taken into account for a precise derivation of the neutrino mass parameters. 

The computation of the NMEs is a subject of debate in the literature for long time, because they bring the large uncertainties 
in the theoretical calculations for DBD. Different groups have developed several conceptually different nuclear structure methods \cite{ROD07}-\cite{RMP10}, 
as we have mentioned in the previous Section. The expression of the NMEs can be written, in general, as a sum of three components:
\begin{equation}
 M^{0 \nu}=M^{0 \nu}_{GT}-\left( \frac{g_V}{g_A} \right)^2 \cdot M^{0 \nu}_F - M^{0 \nu}_T \ ,
\end{equation}
where $M^{0 \nu}_{GT}$, $M^{0 \nu}_F$ and  $M^{0 \nu}_T$ are the Gamow-Teller ($GT$), Fermi($F$) and  Tensor($T$) components, respectively. These are defined as follows:
\begin{equation}
M_\alpha^{0\nu} = \sum_{m,n} \left< 0^+_f\| \tau_{-m} \tau_{-n}O^\alpha_{mn}\|0^+_i \right> \ ,
\end{equation}
where $O^\alpha_{mn}$ are transition operators ($\alpha=GT,F,T$) and the summation is performed over all the nucleon states.
An important part of the NME calculation is the computation of the reduced matrix elements of the two-body transition operators $O^{\alpha}$. Their calculation can be decomposed into products of reduced matrix elements within the spin and relative coordinate spaces. Their explicit expressions are \cite{VES12}, \cite{HS10}:
\begin{equation}
\fl O_{12}^{GT} = \sigma _1 \cdot \sigma _2 H(r) \ , \ \ \ \ \
O_{12}^{F} = H(r) \ , \ \ \ \ \
O_{12}^{T} = \sqrt{\frac{2}{3}} \left [ \sigma _1 \times \sigma _2 \right ]^2 \cdot \frac{r}{R} H(r) C^{(2)}(\hat r) \ .
\end{equation}
The most difficult is the computation of the radial part of the two-body transition operators, which contains the neutrino potentials.  
These potentials depend weakly on the intermediate states and are defined by integrals of momentum carried by the virtual neutrino exchanged between the two nucleons \cite{sim-09}:
\begin{equation}
\fl H_{\alpha} (r) = \frac{2R}{\pi}  \int^\infty_0 j_i (qr) \frac{h_{\alpha}(q)}{\omega} \frac{1}{\omega + \left<E\right>}q^2 dq \equiv \int^\infty_0 j_i (qr) V_{\alpha}(q) q^2 dq \ ,
\label{n_potential}
\end{equation}
where $R=r_0 A^{1/3}$ fm ($r_0=1.2fm$), $\omega = \sqrt{q^2+m_\nu^2}$ is the neutrino energy and $j_i(qr)$ is the spherical Bessel function (i = 0, 0 and 2 for GT, F, and T, respectively).
Usually, in calculations one uses the closure approximation which consists of a replacement of the excitation energies of the states in the intermediate odd-odd nucleus contributing to the decay, by an average expression $\left<E\right>$. 
This approximation works good in the case of $0\nu\beta\beta$ decay modes and simplifies much the calculations. 
The expressions of $h_{\alpha} (\alpha = GT, F, T)$ are:
\begin{equation}
\fl h_F = G_V^2(q^2) \ ,
\end{equation}
\begin{equation}
\fl h_{GT}(q^2) = \frac {G^{2}_A(q^2)}{g^{2}_A} \left [ 1- \frac{2}{3}\frac{q^2}{q^2+m^2_\pi} + \frac{1}{3}\left( \frac{q^2}{q^2+m^2_\pi} \right )^2 \right ]+ \frac{2}{3} \frac {G^{2}_M(q^2)}{g^{2}_A}\frac{q^2}{4m^2_p} \ ,
\label{hgt-hoc}
\end{equation}
and
\begin{equation}
\fl h_{T}(q^2) = \frac {G^{2}_A(q^2)}{g^{2}_A} \left [\frac{2}{3}\frac{q^2}{q^2+m^2_\pi} - \frac{1}{3}\left( \frac{q^2}{q^2+m^2_\pi} \right )^2 \right ]+ \frac{1}{3} \frac {G^{2}_M(q^2)}{g^{2}_A}\frac{q^2}{4m^2_p} \ ,
\label{ht-hoc}
\end{equation}
where $m_\pi$ is the pion mass, $m_p$ is the proton mass and
 \begin{equation}
G_M(q^2) = (\mu_p - \mu_n)G_V(q^2),
\end{equation}
with $(\mu_p - \mu_n)=4.71$.

The expressions (9)-(10) include important nuclear ingredients that should be taken into account for a precise computation of the NMEs, such as the higher order currents in the nuclear interaction (HOC) and finite nucleon size effect (FNS).
Inclusion of HOC brings additional terms in the $H_{GT}$ component and leads to the appearance of the $H_T$ component in the  expressions of the neutrino potentials. 
 FNS effect is taken into account through $G_V$ and $G_A$ form factors:
\begin{equation}
G_A \left(q^2 \right) = g_A \left( \frac{\Lambda^2_A}{\Lambda^2_A+q^2} \right)^2, \ G_V \left( q^2 \right) = g_V \left( \frac{\Lambda^2_V}{\Lambda^2_V+q^2} \right)^2
\label{formfactors}
\end{equation}
For the vector and axial coupling constants the majority of the calculations take either the quenched value, $g_V = 1$ or the 
unquenched one, $g_A = 1.25$, while the values of the vector and axial vectors form factors are $\Lambda_V=850 MeV$ and 
$\Lambda_A=1086 MeV$ \cite{AEE08}, respectively. As one can see, when HOC and FNS corrections are included in the calculations, 
the dependence of NMEs expression on $g_A$ is not trivial and the NMEs values obtained with the quenched or the unquenched value of 
this parameter can not be obtained from by simply re-scaling.

To compute the radial matrix elements $\left<nl|H_{\alpha}|n'l'\right>$ an important ingredient is the adequate inclusion of 
SRCs, induced by the nuclear interaction. The way of introducing the SRC effects has also been subject of debate 
(\cite{KOR07},\cite{sim-09}).  The SRC effects are included by correcting the single particle w. f. as follows:
\begin{equation}
\psi_{nl}(r) \rightarrow \left[ 1+f(r) \right] \psi_{nl}(r) \ .
\end{equation}
The correlation function $f(r)$ can be parametrized in several ways. There are three parameterizations which are used, Miller-Spencer (MS), 
UCOM and CCM (with CD-Bonn and AV18 potentials). The Jastrow prescription \cite{TOM91} for the correlation function is: 
\begin{eqnarray}
f(r) = - c \cdot e^{-ar^2} \left( 1-br^2 \right) \ ,
\label{src}
\end{eqnarray}
\noindent
and includes all these parameterizations, depending of values of the a, b, c parameters. %For example, for $c =0$ and $c=1$ one gets the MS and CCM forms of the correlation function, respectively. 

Including HOC and FNS effects, the radial matrix elements of the neutrino potentials become:
\begin{equation}
\fl \left\langle nl \mid H_\alpha(r) \mid n^\prime l^\prime \right\rangle \ = \ \int^\infty_0 r^2 dr \psi_{nl}(r) \psi_{n^\prime l^\prime}(r)\left[ 1+f(r) \right]^2 \times \int^\infty_0 q^2 dq V_{\alpha} (q) j_n (qr) \ ,
\label{H-two-integrals}
\end{equation}
where $\nu$ is the oscillator constant and $V_{\alpha}(q)$ is an expression containing the q dependence of the neutrino potentials.

From Eqs. (4)-(15) one can see that a set of approximations and parameters are involved in the NMEs expressions, as the HOC and FNS and SRC effects, and $g_A$, $r_0$, ($\Lambda_A$, $\Lambda_B$), $<E>$ parameters. Are there any recommendations on how should they be included in the calculations? At present there is a general consensus in the community in this respect, that will be discussed in the next section.

\section{Numerical results and discussions}

The neutrino mass parameters are derived from Eq. (3). To get $<m_\nu>$ in the same units as $m_e$ we take the NMEs dimensionless and the PSFs ($G^{0\nu}$) in units of $[yr]^{-1}$.

The PSF values were obtained by recalculating them with our code, developed in  \cite{stoica-mirea-prc88}, but with improved numerical precision. 
At this point we mention that the improved PSF values come, on the one hand, by the use of a Coulomb potential describing a more realistic proton charge density in the daughter nucleus instead of the (usual) constant charge density one, 
to solve the Dirac equations for obtaining the electron w.f. On the other hand, we got a better precision of our numerical routines that compute the 
PSFs by enhancing the number of the interpolation points on a case-to-case basis until the results become stationary. 
The obtained values are very close to both those reported previously in refs. \cite{ki-prc85} and \cite{stoica-mirea-prc88}. 
This gives us confidence on their reliability. We mention that these PSFs values differ from older calculations as, for example, those reported in 
refs. \cite{DOI83}, \cite{DOI93}, \cite{SC98} by up to $28\%$. Such differences are important for precise estimations and justify 
the re-actualization of the PSFs values in extracting Majorana neutrino mass parameters.  

For the experimental lifetimes we took the most recent results reported in literature. Especially, we remark the newest results for $^{76}Ge$ from GERDA \cite{ex_gerda} and for $^{136}Xe$ from EXO \cite{ex_exo}.
\newpage
\begin{table}
\caption{The NMEs obtained with different methods. The values are obtained using an unquenched value for $g_A$ and softer SRC 
parametrizations, which are specified in the second column.}
\begin{center}
\begin{tabular}{@{}l@{}l@{}r@{}c@{}c@{}c@{}c@{}c@{}c@{}c@{}c@{}c@{}c@{}} \hline
Method&SRC&\ $^{48}Ca$&\ $^{76}Ge$&\ $^{82}Se$&\ $^{96}Zr$&\ $^{100}Mo$&\ $^{116}Cd$&\ $^{130}Te$&\ $^{136}Xe$&\ $^{150}Nd$ \\ \hline
\cite{ns-jpg41}ShM&CD-BONN	&0.81&3.13&2.88 &	&	&	&	&	&	 \\
\cite{MH12}ShM&CD-BONN		&0.90&	  &	&	&	&	&	&2.21\cite{mh-brown-prl110}	&	 \\
\cite{npa818}ShM&UCOM		&0.85&2.81-3.52&2.64 &	&	&	&2.65	&2.19	&	 \\
\cite{iba-prc2013}IBM-2\ \ &CD-BONN &2.38&6.16&4.99 &3.00	&4.50	&3.29	&4.61	&3.79	&2.88	 \\
\cite{FAE12}QRPA&CD-BONN	&    &5.93(3.27)&5.30(4.54)&2.19	&4.67	&3.72	&4.80	&3.00	&3.16\cite{sim-150nd}	 \\
\cite{jy-qrpa}QRPA&UCOM		&    &5.36(4.11)&3.72 &3.12	&3.93	&4.79	&4.22	&2.80	& 	 \\
\cite{RMP10}GCM&CD-BONN	   	&2.37&4.60&4.22 &5.65	&5.08	&4.72	&5.13	&4.20	&1.71	 \\
\cite{RAH10}PHFB&CD-BONN	&    &	  &	&2.98	&6.07	&	&3.98	&	&2.68	 \\ \hline
  \end{tabular}
 \end{center}
\end{table}

\begin{table}
\begin{center}
\caption[nuclei]{Majorana neutrino mass parameters together with the other components of the $0\nu\beta\beta$ decay halftimes: the $Q_{\beta\beta}$ values, the experimental lifetimes limits, the phase space factors and the nuclear matrix elements.}
\begin{tabular}{@{}r c c c c r@{}}\hline
	  &$Q_{\beta\beta}[MeV]$&\ \ $T^{0\nu\beta\beta}_{exp} [yr]$&\ \ $G^{0\nu\beta\beta} [yr^{-1}]$&$M^{0\nu\beta\beta}$&$\left<m_{\nu}\right>[eV]$\\ \hline
 $^{48}Ca$&4.272&$>5.8\ 10^{22}$\cite{ex_elegant6}	&2.46E-14&0.81-0.90&$<[15.0-16.7]$ \\
 $^{76}Ge$&2.039&$>2.1\ 10^{25}$\cite{ex_gerda}	&2.37E-15&2.81-6.16&$<[0.37-0.82]$\\
	  &					&	 &	    &		   \\
 $^{82}Se$&2.995&$>3.6\ 10^{23}$\cite{ex_nemo3a}	&1.01E-14&2.64-4.99&$<[1.70-3.21]$ \\
 $^{96}Zr$&3.350&$>9.2\ 10^{21}$\cite{ex_nemo3b}	&2.05E-14&2.19-5.65&$<[6.59-17.0]$ \\
$^{100}Mo$&3.034&$>1.1\ 10^{24}$\cite{ex_nemo3a}	&1.57E-14&3.93-6.07&$<[0.64-0.99]$ \\
$^{116}Cd$&2.814&$>1.7\ 10^{23}$\cite{ex_danevich-03}	&1.66E-14&3.29-4.79&$<[2.00-2.92]$\\
$^{130}Te$&2.527&$>2.8\ 10^{24}$\cite{ex_cuoricino}	&1.41E-14&2.65-5.13&$<[0.50-0.97]$ \\
$^{136}Xe$&2.458&$>1.6\ 10^{25}$\cite{ex_exo}		&1.45E-14&2.19-4.20&$<[0.25-0.48]$ \\
$^{150}Nd$&3.371&$>1.8\ 10^{22}$\cite{ex_nemo3c}	&6.19E-14&1.71-3.16&$<[4.84-8.95]$ \\
\hline
\label{tab-ca}
 \end{tabular}
\end{center}
\end{table}
The largest uncertainty in the derivation of $\left<m_{\nu}\right>$ comes from the values of the NMEs. Fortunately, at present there is a general 
consensus in the community on the employment of the different nuclear effects (approximations) and parameters which appear in the 
NMEs expressions (Eqs. (4)-(15)) \cite{giuli-poves-ahep}. Thus, one can restrict the range of spread of the NMEs values for a particular nucleus, 
if one takes into account some recommendations resulting from the analysis of many NMEs calculations. For example, one 
recommends the inclusion in calculation of the HOC, FNS and SRC effects (although their effects can partially compensate 
each other \cite{ns-jpg41}). For SRCs, softer parametrizations like UCOM \cite{KOR07} and CCM \cite{ccm}-\cite{ccm1} are recommended, 
while the MS produces a too sever cut of the w.f. for very short inter-nucleon distances, which reflects into smaller NMEs values. 
Concerning the nuclear parameters, one recommends the use of an unquenched value for the $g_A$ axial vector constant, the values specified above for the vector and axial vector form factors ($\Lambda_V$, $\Lambda_A$), and a value of $r_0=1.2fm$ for the nuclear radius constant. The value for the average energy ($\left<E\right>$), 
used in the closure approximation, is a function of atomic mass A, but the results are less sensitive to changes within a few MeV. The use in different ways 
of these ingredients can result in significant differences between the NMEs values. Hence, a consensus is useful to approach the results obtained by different groups. Having agreement on these nuclear ingredients, the differences in the NMEs values should be searched in the features
of the different nuclear structure methods. These methods use different ways of building the wave functions, different specific model spaces and type of nucleon-nucleon correlations and  use some specific parameters \cite{FAE12},\cite{npa818},\cite{ns-jpg41}. Unfortunately, the uncertainties in the NMEs calculation 
associated with a particular nuclear structure method can not be easily fixed and they are still a subject of debate. As a general feature,  
ShM calculations underestimate the NMEs values (due to the limitations of the model spaces used), while the other methods overestimate them. 
There are, however, a few hints on how to understand/bring closer the NMEs results obtained with different methods. One idea would be to analyze the structure 
of the wave functions used in terms of the seniority scheme \cite{giuli-poves-ahep}. Another one, is to (re)calculate the NME values as to reproduce 
s.p. occupancies numbers measured recently for $^{76}Ge$ and $^{82}Se$ nuclei \cite{SK09}. For example, when the QRPA 
calculations have modified with the s.p. energies that reproduce the experimental occupancies, the new QRPA NMEs values are much closer to the ShM ones.
             
In Table 1 we display the NMEs values obtained with different nuclear methods. For uniformity and in agreement with the consensus discussed above, we chose  those results that were performed with the inclusion of HOC, FNS and SRC(UCOM and CD-Bonn) effects, and with unquenched $g_A=1.25$, as nuclear ingredients. We mention that the newest experimental determinations of this parameter report values even larger (1.269, 1.273) \cite{liu_ga}. However, the differences between NMEs values obtained with $g_A$ = 1.25 - 1.273 are not significant \cite{ns-jpg41}. The NMEs values for $^{76}Ge$ and $^{82}Se$ written in parenthesis represent the adjusted NMEs values obtained with QRPA method by the Tuebingen and Jyvaskyla groups, when the s.p. energies were adjusted to the occupancy numbers reported in ref. \cite{SK09}. One remarks, the QRPA calculations with s.p. occupancies in accordance with experiment, get significantly close to the ShM results, which is remarkable. In the future, one expects measurements of the occupancy numbers for other nuclei, as well. 
Also, it would be interesting if other methods, besides QRPA, would try to recalculate the NMEs by adjusting s.p. energies to experimental occupancy numbers. 

We also make some remarks about the NMEs values on a case-by-case basis.  For $^{48}Ca$ we appreciate that ShM calculations 
give more realistic results than the other methods. In support of this claim we mention, in the case of this isotope, ShM 
calculations are performed within the full pf shell and using good effective NN interactions, checked experimentally on other spectroscopic quantities \cite{ns-jpg41},\cite{MH12}-\cite{poves-adj}. 
Also, we remark that NMEs values obtained with ShM for this isotope were used to correctly predict $T^{2\nu}_{1/2}$, before its experimental measurement \cite{Cau90}.  
For the isotopes with A ={96 - 130} there is a larger spread of the NMEs values calculated with different methods and, consequently, a larger uncertainty in predicting the $\left<m_{\nu}\right>$ parameters. For $^{136}Xe$ there are new ShM large scale calculations with inclusion of a larger model space than the older calculations \cite{mh-brown-prl110}. For this isotope the NMEs values are more grouped. Corroborated with a quite good experimental lifetime, from this isotope one gets the most stringent constraint for the $\left<m_{\nu}\right>$ parameter.

In Table 2 we present our results for the Majorana neutrino mass parameters ($\left<m_{\nu}\right>$) together with the values of $Q_{\beta\beta}$, the 
PSFs ($G^{0\nu}$), NMEs ($M^{0\nu}$) and experimental lifetimes ($T^{0\nu}_{1/2}$) for all the isotopes for which data exists.  Making a sort of the NMEs 
values from literature according to the considerations presented, we reduce the interval of their spread to about a factor of 2, even less (with one exception). This results in reducing the uncertainty in deriving the constraints on the light neutrino Majorana mass parameters, while taking into account NME values obtained with all the main nuclear methods existent on the market. 
One observes that the stringent constraints are obtained from the $^{136}Xe$ isotope, followed by the $^{76}Ge$ one. This is due to both the experimental sensitivity of the experiments measuring these isotopes and to the reliability of the PSFs and NMEs theoretical calculations. The experiments measuring these isotopes are already exploring the quasi-degenerate scenarios for the neutrino mass hierarchy (which around 0.5 eV). With the ingredients presented in Table 2 (PSFs and NMEs) one can appreciate, as well, the sensitivities, translated into neutrino mass parameters, of the future generation of DBD experiments. 
       
\section{Conclusions}

We report new values of light Majorana neutrino mass parameters from a $0\nu\beta\beta$ decay analysis extended to all the isotopes for which theoretical and experimental data exists. We used the most recent results for the experimental lifetimes, $T^{0\nu}_{1/2}$   as well as for the theoretical quantities $G^{0\nu}$ and $M^{0\nu}$. For the PSFs we use newly obtained values, recalculated with an approach described in ref. \cite{stoica-mirea-prc88} but with improved numerical accuracy. We use exact electron w.f. obtained    
by solving a Dirac equation when finite nuclear size and screening effects are included and, in addition, a Coulomb potential derived from a realistic 
proton distribution in the daughter nucleus has been employed. For choosing the NMEs we take advantage on the general consensus in the community on several 
nuclear ingredients involved in the calculations (HOC, FHS and SRCs effects, values of several nuclear input parameters) and restrict the range of spread of 
the NMEs values, reported in the literature. This, in turn, reduces the uncertainty in deriving constraints on the light Majorana neutrino mass parameters, 
while taking into account NME values obtained with all the main nuclear methods. The stringent constraints are obtained from the $^{136}Xe$ and $^{76}Ge$ isotopes, due to both the experimental sensitivity and to the reliability of the PSFs and NMEs calculations. The experiments measuring these isotopes are already exploring the quasi-degenerate scenarios for the neutrino mass hierarchy which is around 0.5 eV.
Our results may be useful for having an up-to-date image on the current neutrino mass sensitivities associated with $0\nu\beta\beta$ measurements for different isotopes and to better estimate the range of the neutrino masses that can be explored in the future DBD experiments.

\section*{Acknowledgments}
This work was done with the support of the MEN and UEFISCDI through the project IDEI-PCE-3-1318, contract Nr. 58/28.10/2011 and Project PN-09-37-01-02/2009.

\end{document}